\documentclass{article}
\usepackage{spconf,amsmath,graphicx,hyperref}
\usepackage{subcaption}			
\usepackage{stfloats} 
\usepackage{multirow} 
\usepackage{booktabs} 
\usepackage{amssymb}
\usepackage{booktabs}
\usepackage{float}
\usepackage{cite}

\title{Unified Multimodal and Multilingual Retrieval via\\ Multi-task Learning with NLU Integration}
\name{
\begin{tabular}{c}
Zhang Xinyuan$^{\star,\dagger}$ \qquad Zhang Lina$^{\star,\dagger}$ \qquad Chen Lisung$^{\star,\dagger}$ \qquad Liu Guangyao$^{\star,\dagger}$ \\
\textit{\qquad Nie Shuai$^{\star}$} \textit{\qquad Xu Jiaming$^{\star}$} \textit{\qquad Shi Runyu$^{\star}$} \textit{\qquad Huang Ying$^{\star}$} \textit{\qquad Zhang Guoquan$^{\star}$}
\end{tabular}
}

\address{$^{\star}$ Xiaomi Corporation, China \\ 
$^{\dagger}$ These authors contributed equally\\
zhangxinyuan8@xiaomi.com}

\begin{document}
%
\maketitle
%
%
\begin{abstract}
Multimodal retrieval systems typically employ Vision Language Models (VLMs) that encode images and text independently into vectors within a shared embedding space. Despite incorporating text encoders, VLMs consistently underperform specialized text models on text-only retrieval tasks. Moreover, introducing additional text encoders increases storage, inference overhead, and exacerbates retrieval inefficiencies, especially in multilingual settings. To address these limitations, we propose a multi-task learning framework that unifies the feature representation across images, long and short texts, and intent-rich queries. To our knowledge, this is the first work to jointly optimize multilingual image retrieval, text retrieval, and natural language understanding (NLU) tasks within a single framework. Our approach integrates image and text retrieval with a shared text encoder that is enhanced by NLU features for intent understanding and retrieval accuracy. 
\end{abstract}
\section{Introduction}
\label{sec:intro}
\subsection{Research Background}
Current retrieval systems \cite{CLIP,maskclip,siglip2} typically require two separate models to process input queries for image and text chunk retrieval. Such as CLIP~\cite{CLIP} adopts a dual-tower architecture in which separate image and text encoders and map inputs to a shared embedding space. In this framework, images are pre-encoded as vectors and stored in a vector database. During inference, only the text encoder is activated to convert the user's queries to vectors and match with the image vector database, which significantly reduces computational overhead. 



In practice, user queries contain both rich information (e.g., time) and noise (e.g., please help me). Directly encoding them may reduce retrieval accuracy. For example, in ``Please help me find a photo of a dog taken last month'', the key information is ``dog", while ``photo" and ``last month" serve as useful contextual constraints to narrow search, and ``Please help me find'' is useless information.Therefore, before image retrieval, the system typically uses a natural language understanding (NLU) model ~\cite{9414110,qiu,Yuan,xu2020} to extract key semantic information from the query, such as intent detection and slot filling, simplifying it into a concise semantic representation suitable for the text encoder.

In addition, in multimodal retrieval systems, users may search text chunks, requiring encoding long texts in documents. Since text encoder in image retrieval systems is mainly optimized for short image descriptions, a separate text encoding model is usually needed for longer content.
This enables the system to retrieve relevant text chunks by aligning the query with semantically similar passages.
Therefore, existing multimodal retrieval systems typically rely on the integration of three components: an image retrieval model, a text retrieval model, and an NLU model. This structure demands significant model storage and computing resources. And it becomes even more challenging to support multiple languages.

There are two solutions for multilingual retrieval systems: one is to use a translator to translate the query into a common language, which introduce significant delays. Another approach is to train separate multilingual models for each task, or multiple small models for different languages. Both methods significantly increase the complexity of a system.
In summary, improving the long text alignment and multilingual processing capabilities of the text encoder in retrieval systems will reduce the system's reliance on multiple models, lower storage and inference costs, and improve search efficiency.

\subsection{Motivation and Contribution}
To address the above challenges in retrieval systems, we propose a multi-task learning method that builds a unified multimodal model for multilingual and multi-task retrieval through phased training. The model extracts search intention from multilingual queries and supports query-to-image and query-to-chunk retrieval across languages. Leveraging multi-task learning, the model optimizes the semantic capability of its text encoder, integrating features across tasks and thereby improving both semantic understanding and generalization.
\section{Method}
\label{sec:method}

\subsection{Model Architecture}
\begin{figure}[t]
  \centering
   \includegraphics[width=1.0\linewidth]{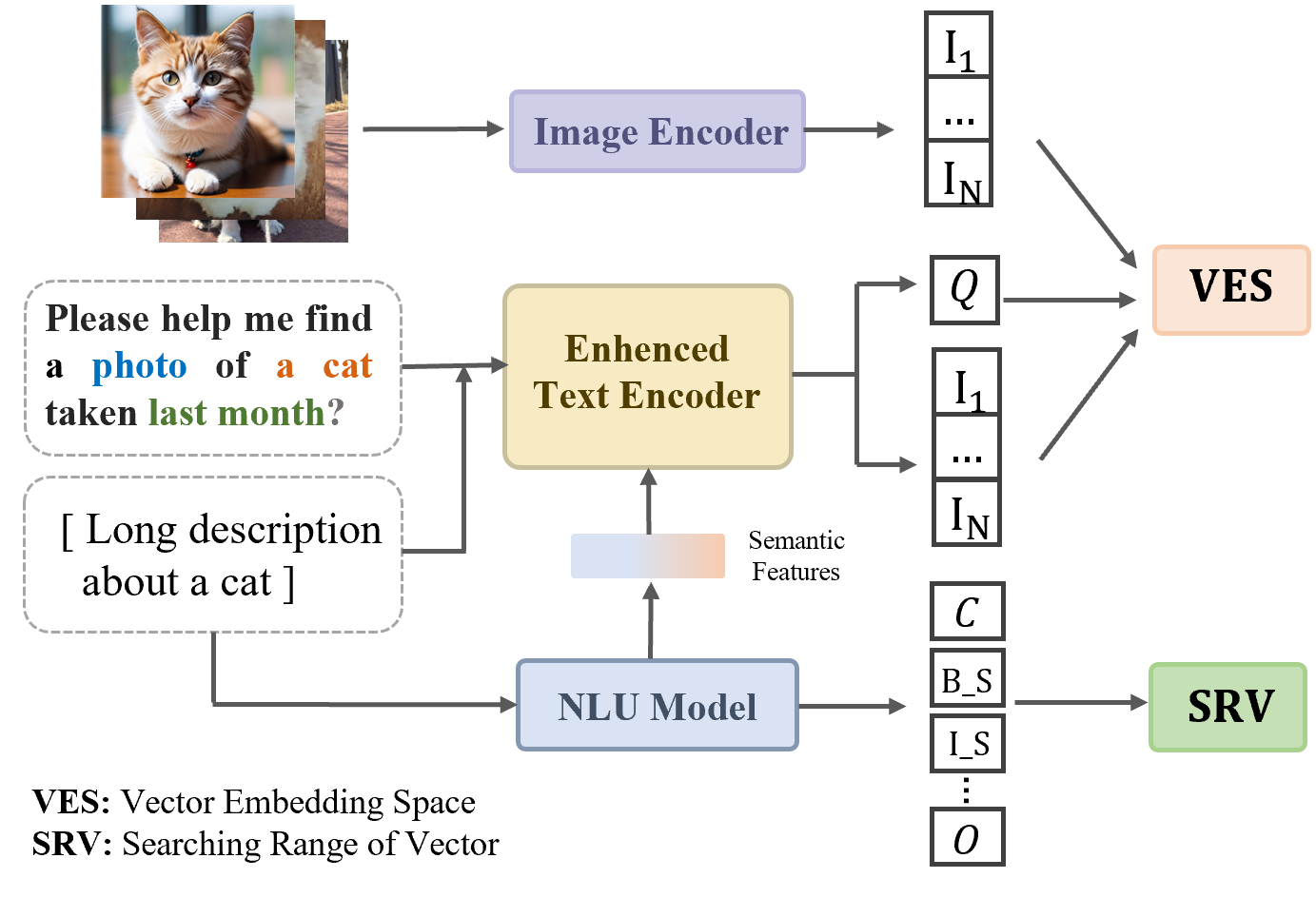}

   \caption{A new multimodal multi-task retrieval system.}
   \label{fig:newarch}
\end{figure}

Fig.~\ref{fig:newarch} shows the designed multimodal multi-task framework, which includes an image encoder, a text encoder, and an NLU module. The framework uses an enhanced text encoder to process query text, short image-describing text, and long fragment text simultaneously. To improve query understanding, NLU semantic features are fed into the text encoder. Although this framework reduces storage and inference costs, it is harder to train than typical frameworks. Therefore, the next section introduces a novel training method for it.

\vspace{-1em}

\subsection{Training Method}
\begin{figure*}[t]
  \centering
   \includegraphics[width=1.0\linewidth]{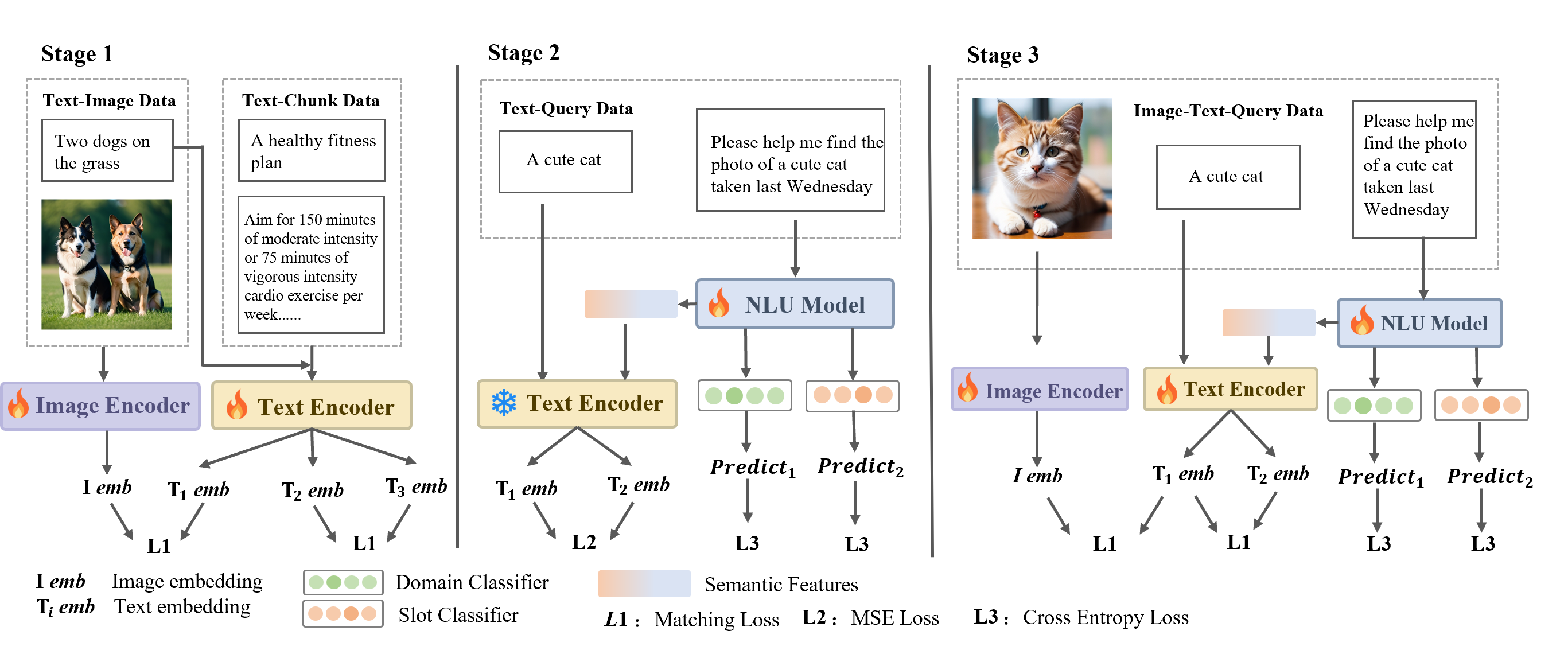}

   \caption{A unique three-stage training method for the proposed framework.}
   \label{fig:fig3}
\end{figure*}

Fig.~\ref{fig:fig3} shows the model training method within our newly designed architecture, which consists of three key stages.


\textbf{Stage 1}: Train the text encoder and finetune the image encoder to equip the model with text-chunk retrieval while preserving text-image alignment functionality.


To reduce costs and improve training efficiency, we selected pre-trained text and image encoders in CLIP~\cite{CLIP} as our base model. During training, we use two types of paired data: text-image pairs and text-chunk pairs. Text-chunk pairs enable self-supervised training of the text encoder to produce similar embeddings for semantically identical texts of varying lengths, while the image encoder is fine-tuned on text-image pairs to maintain its alignment with the text encoder. 

To ensure that the text encoder can effectively encode both descriptive text-image pairs and variable-length texts from text-chunk pairs while sharing weights, we use contrastive learning. The loss function is defined as follows:
\begin{equation}
\small
\mathcal{L}_1 = \mathcal{L}_{TI}+ \alpha \mathcal{L}_{TC}
\end{equation}
where $\alpha$ is a balancing coefficient that adjusts the relative contribution of the query-image and query-chunk losses.

The loss function $L_{TI}$ encompasses both image-to-text loss and text-to-image loss. Thus, the loss calculation for N text-image pairs in a batch is as follows:
\begin{equation}
\scriptsize
{\mathcal{L}_{TI}} =  - \frac{1}{2}{\mathbb{E}_{p(N)}}[\log \frac{{\exp ({f_\theta }(i,t))}}{{\sum\limits_{\tilde t \in T} {\exp } ({f_{\theta }}(i,\tilde t))}} + \log \frac{{\exp ({f_\theta }(i,t))}}{{\sum\limits_{\tilde i \in I} {\exp } ({f_\theta }(\tilde i,t))}}
\end{equation}
Similarly, the loss function $L_{TC}$ also includes text-to-chunk loss and chunk-to-text loss. 


\textbf{Stage 2}: Freeze the parameters of the text encoder trained in Stage 1, and train the NLU module separately to complete intent detection and slot filling tasks.

The input of the NLU module is the query from the user, and the output includes three parts: intent prediction, slot prediction, and a semantic feature vector. The intent detection and slot filling is trained with cross-entropy loss, as follows:

\begin{equation}
\small
\begin{aligned}\mathcal{L}_2=-\sum_{i=1}^Ng^i\log y^i-\sum_{i=1}^N\sum_{j=1}^M\hat{g}_j^i\log\hat{y}_j^i\end{aligned}
\end{equation}
where $g^i$ represents the intent label, $y^i$ represents the predicted intent. $\hat{g}_j^i$ represents the label of the $j$-th slot in the $i$th text, while $\hat{y}_j^i$ represents the predicted slot of the $j$-th token in the $i$-th text.

To incorporate the NLU semantic features into the trained text encoder, we propose the following training steps to enable the NLU module to guide the text encoder using its output features. First, extract the parts of the query labeled as semantic information and merge them into a concise semantic text. Feed this semantic text into the text encoder trained in stage 1 to produce Vector 1. Second, use a cross-attention mechanism to integrate the NLU semantic features with the original query. Then feed this enhanced query representation into the Stage 1 text encoder to obtain Vector 2.

To train the NLU module to effectively guide the text encoder, the similarity between Vector 1 and Vector 2 is supervised using a Mean Squared Error (MSE) loss:
\begin{equation}
\small
\mathcal{L}_3=\frac{1}{N}\sum_{i=1}^N(M_{i}-T_i)^2
\end{equation}
where $N$ is the total number of NLU text pairs in a batch; $M_{i}$ represents the vector generated by encoding the $i$-th text using the text encoder after combining NLU intent features, and $T_i$ represents the vector generated by directly encoding the corresponding ``semantic text'' through the text encoder.
This approach ensures that the text encoder, after integrating NLU features, can capture the semantic intent within the query more accurately.


\textbf{Stage 3}: We jointly finetune the image encoder, text encoder, and NLU module to further improve the model’s overall capabilities in image-text alignment, text matching, and semantic understanding. At this stage, we unfreeze the image encoder and text encoder trained in Stage 1, and the NLU module trained in Stage 2, and jointly finetune them using our multi-task dataset. To support multi-task learning, each training batch includes three types of training pairs.

(1) Image-Text Pair: Consisting of a query and a image, used for text-image alignment.

(2) Short-Long Text Pair: Consisting of a query and a chunk, used for short-to-long text matching.

(3) Intent detection and Slot Filling: Using the query itself to test intent detection and slot filling tasks.


During training, we use a weighted sum of three losses as the overall loss function to balance across the three tasks:

\begin{equation}
\small
\mathcal{L}_{4}=\mathcal{L}_{QI}+a\mathcal{L}_{QC}+b\mathcal{L}_{NLU}
\end{equation}
where $\mathcal{L} _{QI}$ is the matching loss for image-text pairs, ensuring proximity between images and corresponding text in the encoding space.  
$\mathcal{L} _{QC}$ is the matching loss for short-long text pairs, promoting semantic consistency between query and long-form text.  
$\mathcal{L} _{NLU}$ is the loss for intent detection and slot filling, enhancing the model's ability to recognize intent and fill slots in query text.  
The parameters $a$ and $b$ are weights used to balance the contributions of the three tasks. 


\subsection{Data and Training Details}


We adopt LaBSE Vit-L/14~\cite{labse} as the base model, supporting 109 languages, and use a BERT-based NLU~\cite{} module with two prediction heads. In stages 1 and 2, we employ both open-source and self-generated datasets (see Table~\ref{tab:training_data}).

In stage 3, we construct a query-long text-image dataset, named COCO-QLTI, based on MSCOCO and MSCOCO-LONG. Short descriptions from MSCOCO are used to generate queries with intent information via GPT-4o, which are then combined with MSCOCO-LONG’s long description texts and images. The dataset is translated into 12 languages, resulting in 30k pairs per language (360k pairs in total).

All models are trained on 8 A800 GPUs using the AdamW optimizer with a batch size of 1024. The learning rate is initialized at 5e-5 with a warmup ratio of 0.1, and training is conducted for 20, 15, and 20 epochs in Stages 1–3.

\begin{table}[ht]
\centering
\renewcommand{\arraystretch}{1.}
\setlength{\tabcolsep}{3pt}
\small
\caption{Training Datasets across Stages}
\begin{tabular}{l l l c c}
\toprule
Stage & Type & Source & languages & Pairs \\
\midrule
1 & T-I & LAION/WIT~\cite{LAION5B,WIT} & 12 & 200k \\
  & T-I & ShareGPT4~\cite{ShareGTP4}& 12 & 280k \\
  & T-T & xP3/MTP/Miracl~\cite{xP3,MTP,miracl} & 12 & 476k \\
\midrule
2 & NLU & Annotated/Translated & 12 & 202k \\
\midrule
3 & Q-C-I & COCO-QLTI & 12 & 360k \\
\bottomrule
\end{tabular}
\label{tab:training_data}
\end{table}
\section{Evaluation}
\label{sec:eval}
\subsection{Experimental Setup}
We evaluate model performance on both text-image (T2I) and text-text (T2T) retrieval tasks. Table 1 summarizes the model variants and their definitions. Across all results, “Mean” denotes the average performance across languages, and NLU indicates the inclusion of the NLU module.
\begin{table}[ht]
\centering
\small
\caption{Model Variants Used in Ablation Studies.}
\begin{tabular}{lp{5cm}}  
\toprule
\textbf{Model} & \textbf{Training Configuration} \\
\midrule
Ours-stage1 & Stage1 model trained on T2I and T2T  \\
Ours-stage3 & Full three-stage model including NLU \\
LaBSE-FT & Stage1 model trained on T2I  \\
Ours-stage3-w/o NLU & Full 3 stage model without NLU  \\
\bottomrule
\end{tabular}
\label{tab:model-variants}
\end{table}
\vspace{-1em}

\subsection{Image Retrieval Experiment}



In Table~\ref{Performance_combined}, we evaluate the multilingual text-to-image retrieval capabilities of our method on the XTD10~\cite{XDT} and Multi30K* datasets. Compared with SigLIP2~\cite{siglip2}, Jina-Clip-v2~\cite{jinaclip-v2}, and multilingual text-vision encoder baselines
, our method outperforms all baselines on average across multiple languages, surpassing Jina-Clip-v2~\cite{jinaclip-v2} by 1.1\%. 
Results on both datasets further confirm the effectiveness of our model, which achieves strong retrieval performance in most languages.

\begin{table}[h]
\renewcommand{\arraystretch}{1.3}
\centering
\caption{Performance of T2I Retrieval (R@10) }
{\small   
\setlength{\tabcolsep}{3pt}
\renewcommand{\arraystretch}{1.} 
\begin{tabular}{lcc}
\toprule
\textbf{Model} & \textbf{XTD10(Mean)} & \textbf{Multi30K(Mean)} \\ 
\midrule
SigLIP2~\cite{siglip2} & 83.4 & 87.9 \\
Jina-Clip-v2~\cite{jinaclip-v2} & 92.2 & 92.1 \\
LaBSE ViT-L/14~\cite{mclip} & 87.2 & 87.9 \\
XLM-R Large ViT-B/32~\cite{mclip} & 87.9 & 88.3 \\
XLM-R Large ViT-L/14~\cite{mclip} & 89.1 & 90.8 \\
XLM-R Large ViT-B/16+~\cite{mclip} & 91.9 & 92.9 \\
Ours-stage3 & \textbf{93.3} & \textbf{94.8} \\
\bottomrule
\end{tabular}
}
\label{Performance_combined}
\end{table}
\vspace{-1em}

\subsection{Text Retrieval Experiment}
For multilingual text-to-text retrieval, we use COCO-QLTI, which covers 12 languages with 1,000 query–long text pairs each. Baselines include LaBSE~\cite{labse}, Jina-CLIP-v2~\cite{jinaclip-v2}, and M3-Embedding~\cite{BGE}. As shown in Table~\ref{tab:performance_mean_split}, our full model (Ours-stage 3) surpasses LaBSE by up to 48.4$\%$ on average and consistently outperforms all baselines in multilingual mean results. 

\begin{table}[ht]
\renewcommand{\arraystretch}{1.1}
\centering
\caption{Performance on T2T Retrieval (R@10)}
\small
\setlength{\tabcolsep}{6pt}

\begin{tabular}{lcccc}
\cmidrule(lr){1-4}
\textbf{Baselines} & LaBSE~\cite{labse} & M3-Emb~\cite{M3} & Jina-CLIP-v2~\cite{jinaclip-v2} \\
\cmidrule(lr){2-4}
\textbf{Mean} & 35.0  & 79.0 & 82.8 \\
\cmidrule(lr){1-4}
\textbf{Ours} & Ours-stage1 & Ours-s3-w/o NLU & Ours-stage3 & ~ \\
\cmidrule(lr){2-4}
\textbf{Mean} & 72.9 & 80.3 & \textbf{83.4} & ~ \\
\cmidrule(lr){1-4}
\end{tabular}
\label{tab:performance_mean_split}
\end{table}
\vspace{-1em}

\subsection{NLU Experiment}
To assess the effectiveness of jointly training three tasks in stage 3, Table~\ref{Performance of NLU} presents the average performance of the NLU module in 12 languages after stage 2 and stage 3 training. The evaluation focuses on intent detection and slot filling. The NLU module in stage 3 shows better performance than stage 2 on both tasks. Intent detection is a sequence-level classification task, while slot filling is a token-level classification task. The results demonstrate that stage 3 training can improve text representation at different levels of granularity, leading to better performance in both NLU tasks.

\begin{table}[ht] 
\renewcommand{\arraystretch}{1.} 
\centering 
\small
\caption{Performance on NLU at stage 2 and 3} 
\begin{tabular}{lcll} 
\toprule \textbf{Stage} & \textbf{Metric} & \textbf{Mean} \\ 
\midrule \multirow{2}{*}{Stage 2} & Domain Intent (Acc)& 96.35 \\ & Slot (F1) & 88.24 \\
\midrule \multirow{2}{*}{Stage 3} & Domain Intent (Acc)& 96.57 \\ & Slot (F1) & 88.26 \\
\bottomrule \end{tabular}
\label{Performance of NLU}
\end{table}
\vspace{-1em}

\subsection{Ablation Study}
In Stage 1, the model is trained with both text-image and text-text retrieval. To measure the impact of joint training, we compare it with LaBSE-FT, trained only on text-image retrieval. Table~\ref{tab:ablation_stage3} shows that adding the text-text objective improves semantic understanding and text-to-text retrieval while keeping image-to-text performance stable, confirming the benefit of joint training in this stage.
In Stage 3, we integrate an NLU module and jointly optimize text-text retrieval with NLU objectives. Table~\ref{tab:ablation_stage3} compares Ours-Stage 3 with variants and baselines. The results show that Ours-Stage 3 consistently outperforms earlier stages and the variant without NLU, confirming that integrating NLU enriches text representations and yields additional gains in the tri-task framework.
\begin{table}[ht]
\centering
\caption{Ablation results of Stage 1/3 (R@10, Mean)}
\renewcommand{\arraystretch}{1.2}
\setlength{\tabcolsep}{6pt}
\small
\begin{tabular}{lcc}
\toprule
\textbf{Stage 1 Models} & \textbf{XTD10} (T2I) & \textbf{Multi30K} (T2I) \\
\midrule
LaBSE-FT      & 93.3 & 97.1 \\
Ours-stage1   & \textbf{94.4} & \textbf{97.1} \\
\midrule
\textbf{Stage 3 Models} & \textbf{MKQA} (T2T) & \textbf{COCO-QLTI} (T2T) \\
\midrule
LaBSE-FT        & 27.9 & 68.4 \\
Ours-stage1     & 68.2 & 83.4 \\
Ours-stage3-w/o & 68.0 & 82.9 \\
Ours-stage3     & \textbf{69.3} & \textbf{85.1} \\
\bottomrule
\end{tabular}
\label{tab:ablation_stage3}
\end{table}
\vspace{-2em}
\section{Conclusion}
\label{sec:conclu}
This paper introduces a unified framework for natural language search that supports text-image retrieval, text-to-text retrieval, and NLU tasks. To bridge domain gaps, we propose a multi-stage training strategy that jointly optimizes all tasks within a shared embedding space, allowing mutual performance gains across modalities. The resulting model achieves strong multilingual performance, reaching state-of-the-art results on several languages in text-image retrieval, while remaining competitive with prior single-task models on text-to-text retrieval, particularly for NLU-related queries.

\bibliographystyle{IEEEbib}
\bibliography{strings,refs}
\end{document}